\documentclass[12pt,a4paper]{article}

\usepackage{amssymb,amsmath,latexsym,bm,amsfonts}

\usepackage{graphicx}
\usepackage{psfrag,fancyhdr,epsfig}

\usepackage{jheppub}

\usepackage{color}

\usepackage{slashed}
\usepackage{latexsym}

\newcommand{\be}{\begin{equation}}
\newcommand{\ee}{\end{equation}}
\newcommand{\bear}{\begin{array}}
\newcommand{\eear}{\end{array}}

\title{Ultraviolet behaviour of Higgs inflation models}

\author[a, b]{Ignatios Antoniadis}
\author[a]{Anthony Guillen}
\author[c]{Kyriakos Tamvakis}

\affiliation[a]{Laboratoire de Physique Th\'eorique et Hautes Energies - LPTHE\\
Sorbonne Universit\'e, CNRS, 4 Place Jussieu, 75005 Paris, France}
\affiliation[b]{Institute for Theoretical Physics, KU Leuven, Celestijnenlaan 200D,\\ 3001 Leuven, Belgium }
\affiliation[c]{Physics Department, University of Ioannina, 45110, Ioannina, Greece}

\abstract{We study the ultraviolet behaviour of Higgs inflation models above the apparent unitarity violation scale arising from the large non minimal coupling to gravity, by computing on-shell 4-point scattering amplitudes in the presence of a large inflaton background, away from the electroweak vacuum. We find that all tree-level amplitudes are well behaved at high energies below the inflaton background that can thus take values up to the Planck scale. This result holds in both the metric and Palatini formulation, and is independent of the frame (Jordan or Einstein) as expected. The same result also holds if an $R^2$ term is added to the action.}

\begin{document}

\renewcommand{\baselinestretch}{1.2}

\maketitle 

\pagestyle{plain}
\pagenumbering{arabic}


\newpage

\section{Introduction}
Among the existing inflationary models, {\textit{Higgs Inflation}} is particularly attractive since it aims to explain cosmological inflation in terms of a scalar field that plays a central role in conventional particle physics, namely the standard Higgs boson that realises electroweak symmetry breaking at low energies~\cite{BS-1}. The model, although in very good agreement with cosmological observations~\cite{PL}, requires a non-minimal coupling of the inflaton/Higgs to the Ricci scalar $\xi |H|^2R$ with rather large values of the corresponding dimensionless parameter $\xi\sim O(10^4-10^9)$, which through the derivative interactions induced in the Einstein frame seems to cause a tree-level unitarity violation at a scale much below the Planck mass $M_P$~\cite{LM-1}. 

Nevertheless, although it is certain that the non-renormalisable interactions induced by the non-minimal coupling lead to precocious unitarity violation in the electroweak vacuum, the real issue of Higgs Inflation is whether this will show up in the Higgs scattering amplitudes computed {\textit{in the presence of the inflaton background}}. These amplitudes should be computed at energies larger than the alleged unitarity violation scale (deduced on dimensional grounds to be $M_U\,=\,M_P/\xi$ or $M_P/\sqrt{\xi}$ in the case of Palatini variation), in the presence of an inflaton background $\overline{\phi}>>M_P/\sqrt{\xi}$. 

In the above framework, we have calculated the relevant Higgs  tree-level scattering amplitudes in the case of the standard metric formulation of gravity, as well as in the Palatini formulation. We find that the result, although different in the two formulations, is frame-independent, having carried out the computation in both the Jordan and Einstein frames. More importantly, we find that these amplitudes scale with the inverse inflaton background which provides an effective cutoff against unitarity violation up to the Planck scale. This behaviour is common in both formulations, although the theories are different, having different degrees of freedom. 

{{It should be emphasized that our analysis was done for a complex non-minimally coupled scalar field, since for a single non-minimally coupled scalar any apparent unitarity violation can be taken away by a field redefinition~\cite{LM-2}. Earlier estimates~\cite{BS-2} of the unitarity cutoff scale based on a single field model are of limited value not taking into account cancellations at the level of the amplitudes and on-shell conditions. We also note that the UV behaviour of the amplitudes we found is common in both the metric and Palatini formulations. This is in partial disagreement with a recent treatment~\cite{M} of Higgs inflation along similar lines associating this behaviour only with the latter formulation.}}


Our paper is organized as follows. The computation of the relevant 4-point scattering amplitudes at the tree level is performed in Section~2 in the Jordan frame and in Section~3 in the Einstein frame. In Section~4, we repeat the computation in the Palatini formulation, where the connection is treated as independent variable from the metric, while in Section~5, we perform the calculations in the presence of an $R^2$ term for both formulations. {In Section~6, we investigate the possible link between amplitudes in the presence of a large background and in the electroweak vacuum.} Finally, our conclusions are presented in Section~7.

\section{Metric Formulation in the Jordan Frame}

In what follows, concentrating on possible unitarity violation that arises from the non-minimal Higgs coupling, we ignore the potential and consider the action
\be{\cal{S}}=\,\int\,d^4x\,\sqrt{-g}\left\{\frac{1}{2}M_P^2\Omega^2R(g)\,-\,|\partial H|^2\right\}\,,{\label{JM}}\ee 
where
\be \Omega^2\,=\,1+\frac{2\xi}{M_P^2}|H|^2\,.\ee
It suffices to consider the Higgs field as $H=1/\sqrt{2}\left(\phi_1+i\phi_2\right)$ with $\phi_1$ developed around a constant background value $\overline{\phi}_1$:
\be\phi_1\,=\,\phi_1'+\overline{\phi}_1\,\,\,{\textit{with}}\,\,\,\overline{\phi}_1\,\gg\,\frac{M_P}{\sqrt{\xi}}\,.\ee
The metric is developed around a Minkowski background as $g_{ \mu\nu}\,=\,\eta_{ \mu\nu}+\kappa\,h_{ \mu\nu}$. {{Due to the arising mixing between $h_{ \mu\nu}$ and $\phi_1'$, we shall need the expansion of $\sqrt{-g}\,R$ up to third order in $h$, while the expansion of $\sqrt{-g}\,g^{ \mu\nu}$ up to second order will be sufficient. We have:}}
\begin{equation}
\sqrt{-g}R|_h=\kappa(\partial^\mu\partial^\nu h_{\mu\nu}- \Box h),
\end{equation}
\begin{eqnarray}
\sqrt{-g}R|_{h^2}=\frac{\kappa^2}{4}&(&
2h\partial^\mu\partial^\nu h_{\mu\nu}-2h\Box h
-8h^{\mu\nu}\partial_\mu\partial^\rho h_{\nu\rho}+4h^{\mu\nu}\Box h_{\mu\nu} \nonumber\\
&+&4h^{\mu\nu}\partial_\mu\partial_\nu h
-4\partial^\mu h_{\mu\rho}\partial_\nu h^{\nu\rho}
+4\partial^\mu h\partial^\nu h_{\mu\nu}\\
&+&3\partial^\rho h^{\mu\nu}\partial_\rho h_{\mu\nu}
-\partial^\mu h\partial_\mu h 
- 2\partial^\mu h^{\rho\sigma}\partial_\rho h_{\mu\sigma}), \nonumber
\end{eqnarray}
\begin{eqnarray}
\sqrt{-g}R|_{h^3}=\frac{\kappa^3}{8}(
&-&2h^\mu_\nu\partial_\mu h\partial^\nu h
+h\partial_\rho h\partial^\rho h
-4h^\mu_\nu\partial_\rho h^\nu_\mu\partial^\rho h
+4h^\mu_\nu\partial^\sigma h^\nu_\mu\partial_\rho h^\rho_\sigma \nonumber\\
&-&h\partial_\rho h^\mu_\nu\partial^\rho h^\nu_\mu
+4h^\nu_\mu\partial_\nu h^\mu_\sigma\partial^\sigma h
+4h^\mu_\nu\partial^\nu h\partial_\rho h^\rho_\mu \nonumber\\
&-&2h\partial_\rho h^\rho_\sigma\partial^\sigma h
+4h^\mu_\nu\partial_\sigma h^\rho_\mu\partial^\sigma h^\nu_\rho
+2h\partial_\mu h^\rho_\nu\partial^\nu h^\mu_\rho \\
&-&4h^{\mu\nu}\partial_\rho h^\sigma_\mu\partial_\sigma h^\rho_\nu
+2h^\mu_\nu\partial_\mu h^\rho_\sigma\partial^\nu h^\sigma_\rho
-8h^\mu_\nu\partial^\nu h^\sigma_\rho\partial_\sigma h^\rho_\mu),\nonumber
\end{eqnarray}
and
\begin{equation}
\sqrt{-g}g^{\mu\nu}|_h=\frac{\kappa}{2}(h\eta^{\mu\nu}-2h^{\mu\nu}),
\end{equation}
\begin{equation}
\sqrt{-g}g^{\mu\nu}|_{h^2}=\frac{\kappa^2}{8}(h^2\eta^{\mu\nu}-2h^{\rho\sigma}h_{\rho\sigma}h^{\mu\nu}-4hh^{\mu\nu}+8\eta_{\rho\sigma}h^{\mu\rho}h^{\nu\sigma}).
\end{equation}
where $h = h^\mu_\mu$. The second order development $\sqrt{-g}R(g)|_{h^2}$ is often cited after simplifications that use integration by parts. {{Note however that because of the $\Omega^2$ coefficient that depends on position through $\phi_1'$ and $\phi_2$, we will need the above expression in terms of the shifted graviton and before any integration by parts simplifications in order to compute the interaction vertices}}. On the other hand, the third order expansion $\sqrt{-g}R(g)|_{h^3}$ is given after such simplifications.

{\textbf{{Kinetic Terms.}}} The resulting kinetic terms are
\be\bear{l}
{\cal{L}}_{kin\phi_1'}+{\cal{L}}_{kin\phi_2}\,=\,-\frac{1}{2}(\partial\phi_1')^2-\frac{1}{2}(\partial\phi_2)^2\\
\,\\
{\cal{L}}_{mix\phi_1'h}\,=\,\kappa\xi\overline{\phi}_1\phi_1'\left(\partial^{ \mu}\partial^{ \nu}h_{ \mu\nu}-\Box h\right)\\
\,\\
{\cal{L}}_{kin h}\,=\,\frac{1}{4}\kappa^2\xi\overline{\phi}_1^2\left(\frac{1}{2}\partial^{ \mu}h\partial_{ \mu}h-\frac{1}{2}\partial^{ \mu}h^{ \rho\sigma}\partial_{ \mu}h_{ \rho\sigma}-\partial^{ \mu}h\partial^{ \nu}h_{ \mu\nu}+\partial^{ \mu}h^{ \rho\sigma}\partial_{ \rho}h_{ \mu\sigma}\right)\eear\ee
At this point we choose the perturbation factor $\kappa^2=4/(\xi\overline{\phi}_1^2)$. Then, we shift the graviton so that the mixing with $\phi_1'$, after integration by parts, goes away, namely introduce $\tilde{h}_{ \mu\nu}$ as
\be h_{ \mu\nu}\,=\,\tilde{h}_{ \mu\nu}\,-\sqrt{\xi}\phi_1'\eta_{ \mu\nu}\,.\ee 
The resulting scalar kinetic terms are
\be -\frac{1}{2}\left(1+6\xi\right)(\partial\phi_1')^2-\frac{1}{2}(\partial\phi_2)^2\,=\,-\frac{1}{2}(\partial\chi_1)^2-\frac{1}{2}(\partial\phi_2)^2{\label{CANO}}\ee
in terms of the canonically normalised scalar $\chi_1\,\equiv(1+6\xi)^{1/2}\phi_1'$.

Next, in order to obtain the graviton propagator for $\tilde{h}_{ \mu\nu}$ (being the same as for $h_{ \mu\nu}$) we need to introduce gauge-fixing in the form of a term
\be {\cal{L}}_{gf}\,=\,-\left(\partial^{ \mu}\tilde{h}_{ \mu\nu}-\alpha\partial_{ \nu}\tilde{h}\right)^2\,,\ee
{{parametrised in terms of the gauge parameter $\alpha$}}. The resulting ${\cal{L}}_{kin h}+{\cal{L}}_{gf}$ term is
\be -\frac{1}{2}\partial^{ \rho}\tilde{h}_{ \mu\nu}\left(\delta_{\rho}^{\sigma}\delta^{ \mu}_{ \alpha}\delta^{ \nu}_{ \beta}+(2\alpha^2-1)\delta_{ \rho}^{ \sigma}\eta^{\mu\nu}\eta_{ \alpha\beta}-2(2\alpha-1)\delta^{ \sigma}_{ \beta}\eta_{ \rho\alpha}\eta^{ \mu\nu}\right)\partial_{ \sigma}\tilde{h}^{\alpha\beta}\,.\ee
Therefore, the graviton propagator in momentum space is $\frac{-i}{k^2}G_{\alpha\beta}^{\mu\nu}$, where $G_{\alpha\beta}^{\mu\nu}$ is the inverse of the tensor
\be \Delta_{\alpha\beta}^{ \mu\nu}(k)\,=\,\frac{1}{2}\left(\delta_{ \alpha}^{ \mu}\delta_{ \beta}^{ \nu}+\delta_{ \alpha}^{ \nu}\delta_{ \beta}^{ \mu}\right)+2(\alpha^2-1)\eta^{ \mu\nu}\eta_{ \alpha\beta}-(2\alpha-1)\left(\frac{k_{ \alpha}k_{ \beta}}{k^2}\eta^{ \mu\nu}+\frac{k^{ \mu}k^{ \nu}}{k^2}\eta_{ \alpha\beta}\right)\,.\ee
This is inverted to give the graviton propagator
\be -\frac{i}{k^2}G_{ \alpha\beta}^{ \mu\nu}(k)\,=\,-\frac{i}{k^2}\left(I_{ \alpha\beta}^{ \mu\nu}-2T_{ \alpha\beta}^{ \mu\nu}+\frac{(2\alpha-1)}{2(\alpha-1)}C_{ \alpha\beta}^{ \mu\nu}-\frac{(2\alpha-1)^2}{(\alpha-1)^2}K_{ \alpha\beta}^{ \mu\nu}\right)\ee 
in terms of the bi-tensors
\be\bear{l}
I_{ \alpha\beta}^{ \mu\nu}=\frac{1}{2}\left(\delta_{ \alpha}^{ \mu}\delta_{ \beta}^{ \nu}+\delta_{ \alpha}^{ \nu}\delta_{ \beta}^{ \mu}\right),\,\,\,\,\,T_{ \alpha\beta}^{ \mu\nu}\,=\,\frac{1}{4}\eta^{ \mu\nu}\eta_{ \alpha\beta}\\
\,\\
C_{ \alpha\beta}^{ \mu\nu}\,=\,\frac{1}{2}\left(\frac{k_{ \alpha}k_{ \beta}}{k^2}\eta^{ \mu\nu}+\frac{k^{ \mu}k^{ \nu}}{k^2}\eta_{ \alpha\beta}\right),\,\,\,\,K_{ \alpha\beta}^{ \mu\nu}\,=\,\frac{k^{ \mu}k^{ \nu}k_{ \alpha}k_{ \beta}}{(k^2)^2}
\eear   {\label{CT}}\ee

{\textbf{Interactions.}} At this point we decide to focus on the amplitude for $\phi_1,\,\phi_2\,\rightarrow\,\phi_1,\,\phi_2$ scattering. In order to identify the interactions resulting from the $\sqrt{-g}R$ and $\sqrt{-g}g^{ \mu\nu}$ couplings we use the {{developed expressions given previously, including the extra interactions induced from the shifting of the graviton.}} After some calculations it turns out that the only relevant terms in $\sqrt{-g}g^{ \mu\nu}$ are
\be \sqrt{-g}g^{ \mu\nu}\,\supset\,\frac{1}{\xi^{1/2}\overline{\phi}_1}\left(\eta^{ \mu\nu}\tilde{h}-2\tilde{h}^{ \mu\nu}-2\xi^{1/2}\phi_1'\eta^{ \mu\nu}\right)\,,\ee 
while all ${\phi_1'}^2$  terms cancel out. In an analogous fashion, after quite a bit of computation, the other coupling yields the relevant terms:
\begin{equation}	
\sqrt{-g}R|_h=\frac{2}{\xi^{1/2}\bar\phi_1}
(\partial^\mu\partial^\nu\tilde h_{\mu\nu}-\Box\tilde h+3\xi^{1/2}\Box\phi_1')
\end{equation}	
\begin{align}
\sqrt{-g}R|_{h^2} = \frac{4}{\xi^{1/2}\bar\phi_1^2}&\left(
2\partial^\mu\partial^\nu\phi_1'\tilde h_{\mu\nu}
+\partial^\mu\phi_1'\partial^\nu\tilde h_{\mu\nu} \phantom{\frac{1}{2}}\right.\nonumber\\
&\left.-\frac{1}{2}\Box\phi_1'\tilde h + \frac{1}{2}(\partial\phi_1'\cdot\partial\tilde h) 
+ \frac{3}{2}\xi^{1/2}(\partial\phi_1')^2\right)
\end{align}
\begin{align}
\sqrt{-g}R|_{h^3} = -\frac{4}{\xi^{1/2}\overline{\phi}_1^3}&\left(
(\partial^{\mu}\phi_1')(\partial^{\nu}\phi_1')
\left(\tilde{h}_{ \mu\nu}+\frac{1}{2}\eta_{ \mu\nu}\tilde{h}\right)\right.\nonumber\\
&\left.+{\phi_1'}^2(\partial^{\mu}\partial^{\nu}\tilde{h}_{ \mu\nu}-\Box\tilde{h}) 
-3\xi^{1/2}\phi_1'(\partial\phi_1')^2 \phantom{\frac{1}{2}}\right)
\end{align}

Thus, we end up with the following basic interactions, {{expressed in terms of the canonical scalars and the shifted graviton:}}
\be\bear{l}
{\cal{L}}_{\chi_1^3}\,=\,\frac{\chi_1(\partial\chi_1)^2}{\overline{\phi}_1(1+6\xi)^{1/2}},\,\,\,\,{\cal{L}}_{\chi_1^2\phi_2^2}\,=\,\frac{3\xi\phi_2^2(\partial\chi_1)^2}{\overline{\phi}_1^2(1+6\xi)}\\
\,\\
{\cal{L}}_{\chi_1\phi_2^2}\,=\,-\frac{6\xi\phi_2(\partial\chi_1\cdot\partial\phi_2)}{\overline{\phi}_1(1+6\xi)^{1/2}}+\frac{\chi_1(\partial\phi_2)^2}{\overline{\phi}_1(1+6\xi)^{1/2}}\\
\,\\
{\cal{L}}_{\chi_1^2\tilde{h}}\,=\,-\frac{3\xi^{1/2}}{\overline{\phi}_1(1+6\xi)}\chi_1^2\left(\partial^{ \mu}\partial^{ \nu}\tilde{h}_{ \mu\nu}+\Box\tilde{h}\right)\,+\,\frac{(\partial^{ \mu}\chi_1)(\partial^{ \nu}\chi_1)}{\overline{\phi}_1\xi^{1/2}}\left(\tilde{h}_{ \mu\nu}-\frac{1}{2}\eta_{ \mu\nu}\tilde{h}\right)\\
\,\\
{\cal{L}}_{\phi_2^2\tilde{h}}\,=\,\frac{\xi^{1/2}}{\overline{\phi}_1}\phi_2^2\left(\partial^{ \mu}\partial^{ \nu}\tilde{h}_{ \mu\nu}-\Box\tilde{h}\right)+\frac{(\partial^{ \mu}\phi_2)(\partial^{ \nu}\phi_2)}{\overline{\phi}_1\xi^{1/2}}\left(\tilde{h}_{ \mu\nu}-\frac{1}{2}\eta_{ \mu\nu}\tilde{h}\right)
\eear\ee

{\textbf{Calculation of the Jordan frame $\chi_1\phi_2\rightarrow\chi_1\phi_2$ scattering amplitude.}}
The resulting vertices relevant to the $\chi_1,\,\phi_2\,\rightarrow\,\chi_1,\,\phi_2$ scattering amplitude are
\be \bear{l}
V_{\chi_1^2\phi_2^2}\,=\,-\frac{12i\xi}{\overline{\phi}_1^2(1+6\xi)}(p_{\chi_1,1}\cdot p_{\chi_1 2})\\
\,\\
V_{\chi_1^3}\,=\,-\frac{2i}{\overline{\phi}_1^2(1+6\xi)^{1/2}}(p_1\cdot p_2+p_1\cdot p_3+p_2\cdot p_3)\\
\,\\
V_{\chi_1\phi_2^2}\,=\,-\frac{6i\xi}{\overline{\phi}_1(1+6\xi)^{1/2}}p_{\chi_1}^2-\frac{2i}{\overline{\phi}_1(1+6\xi)^{1/2}}(p_{\phi_2 1}\cdot p_{\phi_2 2})\\
\,\\
V_{\chi_1^2\tilde{h}}^{ \mu\nu}\,=\,\frac{6i\xi^{1/2}}{\overline{\phi}_1(1+6\xi)}A^{ \mu\nu}(k_{\tilde{h}})\,-\frac{i}{\overline{\phi}_1\xi^{1/2}}B^{ \mu\nu}(p_{\chi_11}\cdot p_{\chi_1 2})\\
\,\\
V_{\phi_2^2\tilde{h}}^{ \mu\nu}\,=\,-\frac{2i\xi^{1/2}}{\overline{\phi}_1}A^{ \mu\nu}(k_{\tilde{h}})-\frac{i}{\overline{\phi}_1\xi^{1/2}}B^{\mu\nu}(p_{\phi_21}\cdot p_{\phi_2 2})
\eear\ee
where 
\be A_{ \mu\nu}(k)=k_{ \mu}k_{\nu}-\eta_{ \mu\nu}k^2\,\,\,\,{\text{and}}\,\,\,\,\,B^{ \mu\nu}(p_1,p_2)\,=\,p_1^{ \mu}p_2^{ \nu}-\eta^{ \mu\nu}(p_1\cdot p_2)\,,{\label{AB}}\ee
We consider all momenta incoming and have used momentum conservation to simplify expressions.

There are five distinct tree diagrams (and their corresponding amplitudes) generated by these vertices that contribute to $\chi_1,\,\phi_2\rightarrow \chi_1,\,\phi_2$, namely, one from the quartic vertex (${\cal{M}}_4$), two from $\phi_2$-exchange in the $s$ and $u$ channel (${\cal{M}}_{\phi_2, s}$ and ${\cal{M}}_{\phi_2, u}$), and two from $\chi_1$ and $\tilde{h}$ exchange in the $t$-channel (${\cal{M}}_{\chi_1, t}$ and ${\cal{M}}_{\tilde{h}, t}$). We label the momenta according to 
\be \chi_1(p_1),\,\phi_2(p_2)\,\rightarrow\,\chi_1(p_3),\,\phi_2(p_4)\,, \ee 
noting that the momentum of the graviton will be $k=p_1-p_3$. The corresponding Mandelstam variables are $s=(p_1+p_2)^2,\,t=(p_1-p_3)^2$ and $u=(p_1-p_4)^2$.

Using the above vertices, the scalar propagators $i/p^2$ and the graviton propagator $-iG_{\alpha\beta}^{ \mu\nu}/k^2$, we calculate these amplitudes. Nevertheless, it is important to make sure that the amplitudes involving graviton exchange are gauge-invariant (i.e. independent of the gauge-fixing parameter $\alpha$). This is indeed the case based on the fact that contractions of the form $M^{\alpha\beta}L_{\alpha\beta}^{\mu\nu}N_{ \mu\nu}$, where $M$ and $N$ are either $A(k), B(p_1,-p_3)$ or $B(p_2,-p_4)$ {{ (defined in ({\ref{AB}}))}} and $L$ is either $C$ or $K$ {{(defined in ({\ref{CT}}))}}, vanish on shell. The contractions that do not vanish correspond to the $\alpha$-independent terms of the graviton propagator:
$$ A(-k)GA(k)=-\frac{3}{2}t^2,\,B(-p_1,p_3)GB(p_2,-p_4)=\frac{1}{2}(s^2+u^2-t^2)$$
\be A(-k)GB(p_2,-p_4)=-\frac{1}{2}t^2,\,B(-p_1,p_3)GA(k)=-\frac{1}{2}t^2\,.\ee
The resulting amplitudes in the center of mass frame ($s=4E^2,\,t=-2E^2(1-\cos\theta),\,u=-2E^2(1+\cos\theta)$) are
\be \bear{l}
i{\cal{M}}_4\,=\,-\frac{6i\xi}{(1+6\xi)}\frac{t}{\overline{\phi}_1^2}\,,\,\,\,\,\,\,\,i\left({\cal{M}}_{\phi_2, s}+{\cal{M}}_{\phi_2, u}\right)\,=\,\frac{i}{(1+6\xi)}\frac{t}{\overline{\phi}_1^2}\\
\,\\
i{\cal{M}}_{\chi_1, t}\,=\,\frac{it}{\overline{\phi}^2}\,,\,\,\,\,\,i{\cal{M}}_{\tilde{h}, t}\,=\,\frac{i}{\bar\phi_1^2}\left(\frac{12\xi+2}{1+6\xi}t + \frac{1}{2\xi}\frac{s^2+u^2-t^2}{t}\right)
\eear\ee
and the total amplitude is
\be i\mathcal M = -\frac{iE^2}{\bar\phi_1^2}\left(\frac{8(1+3\xi)}{1+6\xi}(1-\cos\theta)+\frac{2(1+\cos\theta)}{\xi(1-\cos\theta)}\right)\,.{\label{JMA}}\ee

\section{Metric Formulation in the Einstein Frame}
Starting from the action ({\ref{JM}}) we convert it to the Einstein frame through a Weyl rescaling of the metric
\be g_{ \mu\nu}\,=\,\Omega^{2}\,\bar{g}_{ \mu\nu}\,,\ee
where 
\be \Omega^2\,=\,1+\frac{2\xi}{M_P^2}|H|^2\,=\,1+\frac{\xi}{M_P^2}\left(\phi_1^2+\phi_2^2\right)\,.\ee
The Einstein frame action is
\be {\cal{S}}\,=\,\int\,d^4x\,\sqrt{-g}\left\{\,\frac{1}{2}M_P^2R(g)\,-\frac{3}{4}\frac{M_P^2}{\Omega^2}\left(\partial\Omega^2\right)^2-\frac{|\partial H|^2}{\Omega^2}\right\}\,\label{action_metric_einstein},\ee 
where we have dropped the {\textit{``bar"}} symbol on the metric for simplicity of notation. 
The scalar part of the action consists of the terms
\be {\cal{L}}_i\,=\,-\frac{1}{2}\frac{(\partial\phi_i)^2}{\Omega^2}-\frac{3\xi^2\phi_i^2}{M_P^2\Omega^4}(\partial\phi_i)^2\,\,\,\,\,{\textit{with}}\,\,\,\,i=1,2\ee 
and
\be {\cal{L}}_{12}\,=\,-\frac{6\xi^2\phi_1\phi_2}{M_P^2\Omega^4}(\partial\phi_1\cdot\partial\phi_2)\,,\ee

Again we consider a constant background value in the $\phi_1$ direction
\be \phi_1\,=\,\phi_1'\,+\,\overline{\phi}_1\,\,\,\,{\textit{with}}\,\,\,\,\overline{\phi}_1>>M_P/\sqrt{\xi}\,.\ee
Then, we have
\be \Omega^{-2}\,\approx\,\frac{M_P^2}{\overline{\phi}_1^2\xi}\left(1-2\frac{\phi_1'}{\overline{\phi}_1}+3\frac{{\phi_1'}^2}{\overline{\phi}_1^2}-\frac{\phi_2^2}{\overline{\phi}_1^2}\right),\,\Omega^{-4}\,\approx\,\frac{M_P^4}{\overline{\phi}_1^4\xi^2}\left(1-4\frac{\phi_1'}{\overline{\phi}_1}+10\frac{{\phi_1'}^2}{\overline{\phi}_1^2}-\frac{2\phi_2^2}{\overline{\phi}_1^2}\right)\,\ee
and
\be {\cal{L}}_{kin \phi_1'}\,=\,-\frac{M_P^2}{\overline{\phi}_1^2}\frac{(1+6\xi)}{2\xi}(\partial\phi_1')^2\,\,\,\,\,{\cal{L}}_{kin\phi_2}\,=\,-\frac{M_P^2}{2\xi\overline{\phi}_1^2}(\partial\phi_2)^2\,.\ee
{{The scalar interaction terms are}}
\be \bear{l}
{\cal{L}}_{{\phi_1'}^3}\,=\,\frac{M_P^2}{\overline{\phi}_1^3}\left(\frac{1+6\xi}{\xi}\right)\phi_1'(\partial\phi_1')^2\\
\,\\
{\cal{L}}_{\phi_1'\phi_2^2}\,=\,\frac{M_P^2}{\xi\overline{\phi}_1^3}\phi_1'(\partial\phi_2)^2-\frac{6M_P^2}{\overline{\phi}_1^3}\phi_2(\partial\phi_1'\cdot\partial\phi_2)\\
\,\\
{\cal{L}}_{{\phi_1'}^2\phi_2^2}\,=\,\frac{M_P^2}{\overline{\phi}_1^4}\left(\frac{1+12\xi}{2\xi}\right)\phi_2^2(\partial\phi_1')^2-\frac{3M_P^2}{2\xi\overline{\phi}_1^4}{\phi_1'}^2(\partial\phi_2)^2+\frac{18M_P^2}{\overline{\phi}_1^4}\phi_1'\phi_2(\partial\phi_1'\cdot\partial\phi_2)
\eear\ee
At this point we introduce the canonically normalised scalars $\chi_1$ and $\chi_2$ as
\be \phi_1'\,=\,\frac{\overline{\phi}_1}{M_P}\left(\frac{\xi}{1+6\xi}\right)^{1/2}\chi_1\,\,\,\,\,{\textit{and}}\,\,\,\,\,\phi_2\,=\,\xi^{1/2}\frac{\overline{\phi}_1}{M_P}\chi_2\,.\ee
In addition to the above interactions of the scalars we also have their interaction to the gravitational perturbation field $g_{ \mu\nu}\approx \eta_{ \mu\nu}+\kappa h_{ \mu\nu}$ through their kinetic terms
\be {\cal{L}}_{\chi_{1,2}^2h}\,=\,\frac{1}{M_P}\left(\partial_{ \mu}\chi_1\partial_{ \nu}\chi_1\,+\,\partial_{ \mu}\chi_2\partial_{ \nu}\chi_2\right)\left(h^{ \mu\nu}-\frac{1}{2}\eta^{ \mu\nu}h\right)\,.\ee

Focusing again on the $\chi_1,\,\chi_2\,\rightarrow\,\chi_1,\,\chi_2$ scattering amplitude we list the relevant vertices. They are
\be\bear{l} \label{vertices_metric_Einstein}
V_{\chi_1^3}\,=\,-\frac{2i}{M_P}\left(\frac{\xi}{1+6\xi}\right)^{1/2}(p_1\cdot p_2+p_1\cdot p_3+p_2\cdot p_3)\\
\,\\
V_{\chi_1\chi_2^2}\,=\,-\frac{2i}{M_P}\left(\frac{\xi}{1+6\xi}\right)^{1/2}(p_{\chi_21}\cdot p_{\chi_22})-\frac{6i\xi}{M_P}\left(\frac{\xi}{1+6\xi}\right)^{1/2}p_{\chi_1}^2\\
\,\\
V_{\chi_i^2,h}^{ \mu\nu}\,=\,-\frac{i}{M_P}\left(p_{\chi_i1}^{ \mu}p_{\chi_i2}^{ \nu}+p_{\chi_i1}^{ \nu}p_{\chi_i2}^{ \mu}-\eta^{ \mu\nu}p_{\chi_i1}\cdot p_{\chi_i2}\,\right)\\
\,\\
V_{\chi_1^2\chi_2^2}\,=\,-\frac{2i\xi}{M_P}\left(\frac{1+12\xi}{1+6\xi}\right)(p_{\chi_11}\cdot p_{\chi_12})+\frac{i}{M_P}\frac{6\xi}{(1+6\xi)}(p_{\chi_21}\cdot p_{\chi_22})\\
\,\\
-\frac{i}{M_P}\frac{18\xi^2}{(1+6\xi)}(p_{\chi_11}+p_{\chi_12})\cdot(p_{\chi_21}+p_{\chi_22})
\eear\ee
The contributing amplitudes are again the one corresponding to the quartic point interaction (${\cal{M}}_4$), two corresponding to a $\chi_2$ exchange in the $s$ and $u$ channels (${\cal{M}}_{\chi_2,s}$ and ${\cal{M}}_{\chi_2,u}$), one corresponding to $\chi_1$-exchange in the $t$-channel (${\cal{M}}_{\chi_1,t}$) and one corresponding to graviton exchange in the $t$-channel (${\cal{M}}_{h, t}$). {{Having carried out gravitational computations in the Jordan frame in an explicitely gauge invariant fashion, we feel confident enough to simplify things by restricting ourselves here to a particular gauge. Thus, using the graviton propagator in the $\alpha=1/2$ gauge}}
\be -\frac{i}{2k^2}G_{ \mu\nu}^{\rho\sigma}(k)\,=\,-\frac{i}{2k^2}\left(\delta_{ \mu}^{ \rho}\delta_{ \nu}^{ \sigma}+\delta_{ \mu}^{ \sigma}\delta_{ \nu}^{ \rho}-\eta_{ \mu\nu}\eta^{ \rho\sigma}\right)\,{\label{HG-PROP}},\ee 
and the scalar propagator $i/p^2$, the above amplitudes are calculated to be
\be\bear{l}
i{\cal{M}}_{\chi_2,s}=-\frac{is}{M_P^2}\frac{\xi}{(1+6\xi)},\,\,\,\,i{\cal{M}}_{\chi_2,u}=-\frac{iu}{M_P^2}\frac{\xi}{(1+6\xi)},\,\,\,i{\cal{M}}_{\chi_1,t}=\frac{i\xi t}{M_P^2}\\
\,\\
i{\cal{M}}_4=\frac{it}{M_P^2}\frac{2\xi(1+3\xi)}{(1+6\xi)},\,\,\,i{\cal{M}}_{h, t}\,=\,-\frac{i}{2M_P^2}\left(1-\frac{s^2+u^2}{t}\right)
\eear\ee
Considering scattering in the center of mass, we have $s = 4\tilde E^2, t=-2\tilde E^2(1-\cos\theta)$ and $u = -2\tilde E^2(1+\cos\theta)$, where $\tilde{E}$ is the energy in the Einstein frame, related to the energy in the Jordan frame as $\tilde E\,\approx\,\frac{M_P}{\overline{\phi}_1\sqrt{\xi}}E$. The resulting overall amplitude is
\be i\mathcal M = -\frac{iE^2}{\bar\phi_1^2}\left(\frac{8(1+3\xi)}{1+6\xi}(1-\cos\theta)+\frac{2(1+\cos\theta)}{\xi(1-\cos\theta)}\right)\,,\ee 
exactly the same as the amplitude ({\ref{JMA}}), calculated in the Jordan frame. 

As before, this amplitude scales like $E^2/\bar\phi_1^2$ when $\xi$ is large. {{In contrast, a scaling $\xi E^2/\bar\phi_1^2 \sim \xi^2\tilde E^2/M_p^2$ requires the existence of an amplitude that scales like $\xi^2$ in the Einstein frame, due to the factor $\xi$ in the relation between the two frames. Nevertheless, as we can see from the vertices (\ref{vertices_metric_Einstein}), the only way to obtain an amplitude that scales as $\xi^2$ in the Einstein frame is to use the $\chi_1\chi_2^2$ vertex two times, due to its second term.}} This has been done when computing $\mathcal{M}_{\chi_2, u}$ and $\mathcal{M}_{\chi_2, u}$, but in these two cases, the $\chi_1$ line attached to the vertex is external, so $p_{\chi_1}^2$ is equal to zero and the dangerous contribution vanishes. Another possibility would be to consider $\chi_2, \chi_2 \rightarrow \chi_2, \chi_2$ scattering. In this case, the $\chi_1\chi_2^2$ can arise two times with an internal $\chi_1$ line. But this contribution would arise in all of the $s, t$ and $u$ channels, with each amplitude being proportional to the corresponding Mandelstam variable. So in the end, we get an expression proportional to $s+t+u = 0$ since the scalars we consider are massless. In fact, when considering $\chi_1, \chi_1 \rightarrow \chi_1, \chi_1$ or $\chi_2, \chi_2 \rightarrow \chi_2, \chi_2$ scattering, the only non vanishing amplitudes are the ones due to graviton exchange:
\begin{equation}
i\mathcal{M}_{\tilde h} = \frac{i}{2M_p^2}\left(\frac{t^2+u^2}{s}+\frac{u^2+s^2}{t}+\frac{s^2+t^2}{u}\right)
\end{equation}
leading in terms of $E$ and $\theta$:
\begin{equation}
i\mathcal{M}_{\tilde h} = -\frac{iE}{\bar\phi_1^2\xi}\frac{(3+\cos^2\theta)^2}{1-\cos^2\theta}
\end{equation}
Similarly, the $2 \rightarrow 2$ amplitudes with external gravitons cannot scale as $\xi^2$ in the Einstein frame. Once the canonically normalised scalars are introduced, the three-point interactions $\chi_i^2 h$ have no $\xi$ dependence. There are four-point interactions $\chi_i^3 h$, but none of them scale as $\xi^2$ in the Einstein frame.

\section{Palatini Formulation}
In the so-called {\textit{Palatini}} or {\textit{first order formulation}}~\cite{P} of GR the metric $g_{ \mu\nu}$ and the connection $\Gamma_{ \mu\nu}^{ \rho}$ are treated as independent variables and the standard Levi-Civita relation relating the two arises as an equation of motion. Although the two formulations are entirely equivalent within GR, in the case of scalar fields non-minimally coupled to gravity the two formulations differ leading to different results~\cite{BD}. In this section we shall analyse the unitarity issue of Higgs inflation in the framework of the Palatini variation. We start again with the action
\be {\cal{S}}\,=\,\int\,d^4x\,\sqrt{-g}\left\{\frac{1}{2}\Omega^2R(g,\omega)-|\partial H|^2\,\right\}\,,\ee 
where the Ricci scalar is a function of the independent metric $g_{ \mu\nu}$ and spin connection $\omega_{a b}^{ \mu}$. Again, $\Omega^2=1+\frac{2\xi}{M_P^2}|H|^2
$. 

Varying the action with respect to $\omega_{a b}^{ \mu}$ we obtain
$$\delta{\cal{S}}=\frac{1}{2}M_P^2\int\,d^4x\,e_a^{ \mu}e_b^{ \nu}\Omega^2\left(D_{ \mu}\delta\omega_{ \nu}^{\,ab}-D_{ \nu}\delta\omega_{ \mu}^{\,ab}\right)$$
\be=\frac{1}{2}M_P^2\int\,d^4x\,e_a^{ \mu}e_b^{ \nu}\Omega^2\left(2\nabla_{ \mu}\delta_{ \nu}^{\,ab}+T_{ \mu\nu}^{\rho}\delta\omega_{ \rho}^{\,ab}\right)\,,\ee
having introduced the torsion. Integrating by parts we arrive at
\be \delta{\cal{S}}\,=\,\frac{1}{2}M_P^2\int\,d^4x\,e\left\{\Omega^2\left(T_{ab}^{\,\nu}+T_{ \rho a}^{\,\rho}e_b^{ \nu}-T_{ \sigma b}^{ \sigma}e_a^{ \nu}\right)-\left(\partial_{a}\Omega^2e_b^{ \nu}-\partial_b\Omega^2e_a^{ \nu}\right)\,\right\}\delta\omega_{ \nu}^{\,ab}\,,\ee
which gives the equation
\be \Omega^2\left(T_{ab}^{\,\nu}+T_{ \rho a}^{\,\rho}e_b^{ \nu}-T_{ \sigma b}^{ \sigma}e_a^{ \nu}\right)\,=\,\partial_{a}\Omega^2e_b^{ \nu}-\partial_b\Omega^2e_a^{ \nu}\,.\ee
This equation is solved to give the torsion
\be T_{ab}^{\,\nu}\,=\,-\frac{1}{2}\Omega^{-2}\left(\partial_a\Omega^2\,e_b^{ \nu}-\partial_b\Omega^2\,e_a^{ \nu}\right)\,.\ee
Next, we can use this to express the contorsion tensor $K_{ \mu\nu\rho}=-\frac{1}{2}(T_{ \mu\nu\rho}-T_{ \nu\mu\rho}+T_{ \rho\mu\nu})=T_{ \nu\rho\mu}$ and insert it in the spin connection, written as a sum of the torsion-free part $\omega_{ \mu}^{\,ab}(e)=2e^{\nu\left[a\right.}\partial_{ \left[\mu\right.}e_{\left.\nu\right]}^{\left.\,b\right]}-e^{\nu\left[a\right.}e^{\left.b\right]\sigma}e_{ \mu c}\partial_{ \nu}e_{\,\sigma}^c$ and $K_{ \mu\nu\rho}$ as
\be \omega_{ \mu\nu\rho}=\omega_{ \mu\nu\rho}(e)+K_{ \mu\nu\rho}\,.\ee
This expression gives the spin connection in terms of $e$ and $\Omega^2(H)$. Inserting it into the expression for $R(e,\omega)=e^{\mu}_ae^{ \nu}_b(\partial_{ \mu}\omega_{ \mu}^{\,ab}-\partial_{ \nu}\omega_{ \mu}^{\,ab}+\omega_{ \mu}^{\,ac}\omega_{ \nu c}^{\,b}-\omega_{ \nu}^{\,ac}\omega_{ \mu c}^{\,b})$ we finally obtain
\be R\,=\,R(g)+2\nabla_{ \mu}K_{ \nu}^{\,\mu\nu}+K_{\mu}^{\,\mu\sigma}K_{ \nu\sigma}^{\,\nu}-K_{ \nu}^{\,\mu\sigma}K_{ \mu\sigma}^{\,\,\nu}\,,\ee 
where $R(g)$ is the standard metric Ricci scalar. Considering now the $K$-part of the action we have
\be \int d^4x\sqrt{-g}\Omega^2\nabla_{ \mu}K_{ \nu}^{\,\mu\nu}=-\int d^4x\sqrt{-g}\partial_{ \mu}\Omega^2\,K_{ \nu}^{\,\mu\nu}=-3\int d^4x\sqrt{-g}\Omega^{-2}(\partial\Omega^2)^2,\ee
while 
\be K_{\mu}^{\,\mu\sigma}K_{ \nu\sigma}^{\,\nu}-K_{ \nu}^{\,\mu\sigma}K_{ \mu\sigma}^{\,\,\nu}\,=\,-\frac{3}{2}\Omega^{-4}(\partial\Omega^2)^2\,.\ee 
Thus, finally, we have the following Palatini form of the action
\be {\cal{S}}\,=\,\int\,d^4x\,\sqrt{-g}\left\{\frac{1}{2}M_P^2\Omega^2R(g)\,+\,\frac{3}{4}M_P^2\frac{(\partial\Omega^2)^2}{\Omega^2}\,-|\partial H|^2\,.\right\}{\label{JP}}\ee
Note that the second term in the action is up to a sign equal to the term that would arise in the metric formulation if we were to make a Weyl rescaling that would take us to the Einstein frame. In fact, if we make a Weyl rescaling $g_{ \mu\nu}=\Omega^{2}\bar{g}_{ \mu\nu}$ in ({\ref{JP}}) the extra term generated will exactly cancel out this term and we will obtain a standard Einstein frame Palatini action of the form $M_P^2\int d^4x\sqrt{-\bar{g}}\left\{R(\bar{g})/2-\Omega^{-2}|\partial H|^2\right\}$. Thus, the appearance of such a term should have been anticipated.

We shall now proceed to eventually analyse the general behaviour of scattering amplitudes calculated on the basis of ({\ref{JP}}) as we did in the case of the metric formulation. Staying in the Jordan frame we introduce a Higgs/inflaton background $\overline{\phi}_1>>M_P/\sqrt{\xi}$ as $\phi_1=\phi_1'+\overline{\phi}_1$. Again we have
\be \,\Omega^{-2}\,\approx\,
\frac{M_P^2}{\overline{\phi}_1^2\xi}\left(1-2\frac{\phi_1'}{\overline{\phi}_1}+3\frac{{\phi_1'}^2}{\overline{\phi}_1^2}-\frac{\phi_2^2}{\overline{\phi}_1^2}\right)\,.\ee 
From here on the analysis is entirely analogous to the analysis carried out in the metric case with the additional contributions of the extra term. The resulting contributions that are relevant to $\phi_1,\,\phi_2\rightarrow\,\phi_1,\,\phi_2$ scattering are (denoting with the superscripts (P) and (M) the corresponding Palatini and metric contributions)
\be \bear{l}
{\cal{L}}_{kin \phi_1'}^{(P)}\,=\,{\cal{L}}_{kin \phi_1'}^{(M)}+3\xi(\partial\phi_1')^2\,=\,-\frac{1}{2}(\partial\phi_1')^2\\
\,\\
{\cal{L}}_{{\phi_1'}^3}^{(P)}\,=\,{\cal{L}}_{{\phi_1'}^3}^{(M)}-\frac{6\xi}{\overline{\phi}_1}\phi_1'(\partial\phi_1')^2\,=\,\frac{\phi_1'}{\overline{\phi}_1}(\partial\phi_1')^2\\
\,\\
{\cal{L}}_{{\phi_1'}\phi_2^2}^{(P)}\,=\,{\cal{L}}_{{\phi_1'}\phi_2^2}^{(M)}\,+\,\frac{6\xi}{\overline{\phi}_1}\phi_1'(\partial\phi_1'\cdot\partial\phi_2)\,=\,\frac{\phi_1'}{\overline{\phi}_1}(\partial\phi_1'\cdot\partial\phi_2)\\
\,\\
{\cal{L}}_{{\phi_1'}^2\phi_2^2}^{(P)}\,=\,{\cal{L}}_{{\phi_1'}^2\phi_2^2}^{(M)}\,
-\frac{18\xi}{\overline{\phi}_1^2}\phi_1'\phi_2(\partial\phi_1'\cdot\partial\phi_2)-\frac{3\xi}{\overline{\phi}_1^2}\phi_2^2\partial\phi_1')^2\,=\,-\frac{18\xi}{\overline{\phi}_1^2}\phi_1'\phi_2(\partial\phi_1'\cdot\partial\phi_2)\\
\,\\
{\cal{L}}_{{\phi_1'}^2\tilde{h}}^{(P)}\,=\,{\cal{L}}_{{\phi_1'}^2\tilde{h}}^{(M)}\,-\frac{3\xi^{1/2}}{\overline{\phi}_1}(\partial^{ \mu}\phi_1')(\partial^{ \nu}\phi_1')\left(\tilde{h}_{ \mu\nu}-\frac{1}{2}\eta_{ \mu\nu}\tilde{h}\right)\\
\,\\
\,=\,-\frac{3\xi^{1/2}}{\overline{\phi}_1}{\phi_1'}^2\left(\partial^{ \mu}\partial^{ \nu}\tilde{h}_{ \mu\nu}\,-\Box\tilde{h}\right)\,+\,\frac{1}{\xi^{1/2}\overline{\phi}_1}(\partial^{ \mu}\phi_1')(\partial^{ \nu}\phi_1')\left(\tilde{h}_{ \mu\nu}-\frac{1}{2}\eta_{ \mu\nu}\tilde{h}\right)
\eear\ee
Note that contrary to the metric case $\phi_1'$ is already canonically normalised and there is no need to introduce new scalars. The resulting interaction vertices are
\be \bear{l}
V_{ \phi_1'\phi_2^2}=-\frac{2i}{\overline{\phi}_1}(p_{\phi_21}\cdot p_{\phi_22})\,,\,\,\,\,\,\,V_{{\phi_1'}^2\phi_2^2}\,=\,\frac{6i\xi}{\overline{\phi}_1^2}(p_{\phi_1'1}+p_{\phi_1'2})\cdot(p_{\phi_21}+p_{\phi_22})\\
\,\\
V_{{\phi_1'}^3}\,=\,-\frac{2i}{\overline{\phi}_1}\left(p_{\phi_1'1}\cdot p_{\phi_1'2}\,+\,p_{\phi_1'1}\cdot p_{\phi_1'1}\cdot p_{\phi_1'3}\,+\,p_{\phi_1'2}\cdot p_{\phi_1'3}\right)\\
\,\\
V_{{\phi_1'}^2\tilde{h}}^{ \mu\nu}\,=\,\frac{6i\xi^{1/2}}{\overline{\phi}_1}A^{ \mu\nu}(k_{\tilde{h}})\,-\frac{i}{\xi^{1/2}\overline{\phi}_1}B^{ \mu\nu}(p_{\phi_1'1},\,p_{\phi_1'2})\\
\,\\
V_{\phi_2^2\tilde{h}}^{ \mu\nu}\,=\,-\frac{2i\xi^{1/2}}{\overline{\phi}_1}A^{ \mu\nu}(k_{\tilde{h}})-\frac{i}{\xi^{1/2}\overline{\phi}_1}B^{ \mu\nu}(p_{\phi_21},\,p_{\phi_22})
\eear\ee
where all momenta are considered incoming and 
\be A_{\mu\nu}(k)=k_{ \mu}k_{ \nu}-k^2\eta_{ \mu\nu}\,\,\,\,{\textit{and}}\,\,\,\,B^{ \mu\nu}(p_1,p_2)\,=\,p_1^{ \mu}p_2^{ \nu}+p_1^{ \nu}p_2^{ \mu}-\eta^{ \mu\nu}(p_1\cdot p_2)\,.\ee

Using these interaction vertices, the scalar propagator $i/p^2$ and {{the $\alpha=1/2$ graviton propagator}} ({\ref{HG-PROP}}) we proceed to calculate the amplitudes corresponding to the relevant tree diagrams that contribute to $\phi_1',\,\phi_2\rightarrow \phi_1',\,\phi_2$ scattering. As previously, we have the quartic contact interaction (${\cal{M}}_4$), $\phi_2$-exchange in the $s$ and $u$ channels (${\cal{M}}_{\phi_2,s}$ and ${\cal{M}}_{\phi_2,u}$), $\phi_1'$-exchange in the $t$ channel (${\cal{M}}_{\phi_1',t}$) and graviton exchange in the $t$ channel (${\cal{M}}_{\tilde{h},t}$). Labelling the momenta as 
$\phi_1'(p_1)\phi_2(p_2)\rightarrow \phi_1'(p_3)\phi_2(p_4)$ and expressing the amplitudes in terms of Mandelstam variables $s=(p_1+p_2)^2,\,t=(p_1-p_3)^2,\,u=(p_1-p_4)^2$, we obtain the amplitudes
\be\bear{l}
i{\cal{M}}_4\,=\,-\frac{18i\xi t}{\overline{\phi}_1^2},\,\,\,\,i\left({\cal{M}}_{\phi_2,s}+{\cal{M}}_{\phi_2,u}\right)\,=\,\frac{it}{\overline{\phi}_1^2},\,\,\,i{\cal{M}}_{\phi_1',t}\,=\,\frac{it}{\overline{\phi}_1^2}\\
\,\\
i{\cal{M}}_{\tilde{h},t}\,=\,\frac{i}{\overline{\phi}_1^2}\left((18\xi+2)t+\frac{s^2+u^2-t^2}{2\xi t}\right)
\eear\ee
In the center of mass frame, denoting by $\theta$ the angle between ingoing and outgoing $\phi_1'$, we have $t=-2E^2(1-\cos\theta)$, $u=-2E^2(1+\cos\theta)$ and $s=4E^2$. Thus, the overall Jordan frame scattering amplitude is
\be i{\cal{M}}\,=\,-\frac{iE^2}{\overline{\phi}^2}\left(8(1-\cos\theta)+\frac{2(1+\cos\theta)}{\xi(1-\cos\theta)}\right)\,.{\label{PAL}}\ee
It has been also checked that the same calculation carried out in the Einstein frame yields exactly the same result. Both amplitudes, in the metric and Palatini formalism, scale in a similar way, proportionally to $E^2/\overline{\phi}_1^2$ when $\xi$ is large. But this time, the absence of a factor $\xi$ in this scaling seems less trivial, since it is due to the cancellation of $\xi$-proportional terms between $\mathcal{M}_4$ and $\mathcal{M}_{\tilde h, t}$. However, doing the calculation in the Einstein frame yields $3$-point vertices scaling like $\sqrt{\xi}$ and $4$-point vertex scaling like $\xi$, from which it is clear that the $2\rightarrow2$ scattering amplitudes cannot scale like $\xi^2$ or higher. With the factor $1/\xi$ that appears when replacing the Einstein frame energy by the Jordan frame energy, this means that $2\rightarrow2$ scattering amplitudes scale proportionally to at most $E^2/\bar\phi_1^2$.

\section{Inflation in the presence of an $\bf R^2$ term}

As it was mentioned in the introduction a most attractive feature of Higgs inflation is the fact that cosmological inflation is directly related to particle physics, the role of the inflaton being taken up by the Higgs field itself. There is an equally appealing philosophy {{ associating inflation directly to gravity, realised in the Starobinsky model~\cite{S},}} which features a quadratic Ricci scalar curvature term in the action, the role of the inflaton being taken up by the extra scalar mode included in the gravitation multiplet. The Starobinsky model of inflation is in good agreement with existing cosmological data~\cite{PL}. Appart from the simple Starobinsky model a number of models have also been proposed featuring both a Higgs-like scalar field, non-minimally coupled to gravity, in the presence of an $R^2$ term~\cite{A}. It is worthwhile to discuss the behaviour of inflaton scattering amplitudes in this class of models as well.

Starting with the very simple case of the Starobinsky model in the metric formulation,\footnote{Note that in the Palatini variation of the Starobinsky model there is no propagating scalar degree of freedom~\cite{A}.} the action can be expressed in terms of the auxiliary scalar $\chi$ as
\be {\cal{S}}=\int\,d^4x\sqrt{-g}\left\{\frac{1}{2}M_P^2R+\frac{\alpha}{4}R^2
\right\}\,=\,\int\,d^4x\sqrt{-g}\left\{\frac{1}{2}M_P^2\Omega^2R-\frac{\alpha}{4}\chi^4\right\}\ee 
with $\Omega^2=1+\frac{\alpha}{M_P^2}\chi^2$. Going to the Einstein frame by a Weyl rescaling $g_{ \mu\nu}\rightarrow\Omega^{-2}{g}_{\mu\nu}$, we obtain the action in the form
\be {\cal{S}}\,=\,\int\,d^4x\sqrt{-g}\left\{\frac{1}{2}M_P^2R-\frac{3M_P^2}{4\Omega^4}(\partial\Omega^2)^2-\frac{\alpha\chi^4}{4\Omega^4}\right\}\,.\ee 
Introducing a canonically normalised scalar $\rho$ through
\be d\rho\,=\,\frac{\alpha\sqrt{6}}{M_P}\frac{\chi\,d\chi}{1+\frac{\alpha}{M_P^2}\chi^2}\,\Longrightarrow\,\bear{l}
\rho = \frac{\sqrt{6}M_P}{2}\ln\left(1+\frac{\alpha\chi^2}{M_P^2}\right)\\
\,\\
\chi^2 = \frac{M_P^2}{\alpha}\left(\exp\left(\frac{2\rho}{\sqrt{6}M_P}\right) - 1\right)
\eear
\ee
we have a canonical scalar $\rho$, minimally coupled to gravity, with a potential
\be V(\rho) = \frac{M_P^4}{4\alpha}\frac{\left(\exp\left(\frac{2\rho}{\sqrt{6}M_P}\right)-1\right)^2}{\exp\left(\frac{4\rho}{\sqrt{6}M_P}\right)}.\ee
This potential exhibits an inflationary plateau in the region $2\rho>>\sqrt{6}M_P$. Setting $\rho=\rho'+\overline{\rho}$, where $2\overline{\rho}>>\sqrt{6}M_P$ (this corresponds to a $\chi$-background $\overline{\chi}>>M_P/\sqrt{\alpha}$), we see that the potential is flat and there is no $\rho'$-mass or any ${\rho'}^3,\,{\rho'}^4$ interactions. Therefore, the only interaction is the minimal gravitational interaction ${\rho'}^2h$. Computing the corresponding amplitude for inflaton scattering\footnote{This is exactly the same for $\chi_1\,\chi_1\rightarrow\chi_1\chi_1$ through $V_{\chi_1^2h}$ in the Einstein frame.} $\rho'\,\rho'\,\rightarrow\,\rho'\,\rho'$, we obtain
\be i{\cal{M}}\,=\,-\frac{iE^2}{\alpha\overline{\chi}^2}\frac{(3+\cos^2\theta)^2}{(1-\cos^2\theta)}\,,\ee
where $E$ is the Jordan-frame energy and $\theta$ is the incoming/outgoing $\rho'$ angle. 
Note that $\alpha\overline{\chi}^2>>M_P^2$. 

\subsection{Palatini Higgs Inflation in the presence of an $\bf{R^2}$ term.}
Moving next to the more involved case of Higgs inflation in the presence of an $R^2$ term, we start with the action
\be {\cal{S}}\,=\,\int\,d^4x\,\sqrt{-g}\left\{\frac{1}{2}\left(M_P^2+2\xi|H|^2\right)R-|\partial H|^2\,+\,\frac{\alpha}{4}R^2\,\right\}\,,{\label{ACTI}}\ee 
with $R=R(g,\omega)$, aiming at analysing scattering in the Einstein frame in the framework of the Palatini variation\footnote{{Having checked explicitly the frame-independence of the calculated scattering amplitudes in our analysis of Higgs inflation, we restrict our study of Palatini-$R^2$ inflation in the simpler Einstein frame.}}. The action can be set in the auxiliary scalar form as
\be {\cal{S}}\,=\,\int\,d^4x\sqrt{-g}\left\{\frac{1}{2}M_P^2\Omega^2R-|\partial H|^2-\frac{\alpha}{4}\chi^4\,\right\}\,,\ee
where
\be \Omega^2\,=\,1+\frac{2\xi}{M_P^2}|H|^2+\frac{\alpha}{M_P^2}\chi^2\,.\ee
Going to the Einstein frame through a rescaling $g_{ \mu\nu}\rightarrow \Omega^{-2}g_{ \mu\nu}$, we obtain
\be {\cal{S}}\,=\,\int\,d^4x\sqrt{-g}\left\{\frac{1}{2}M_P^2R-\frac{1}{\Omega^2}|\partial H|^2-\frac{\alpha\chi^4}{4\Omega^4}\,\right\}\,.\ee 
Variation with respect to the auxiliary $\chi$ gives an algebraic equation with the solution
\be \chi^2\,=\,\frac{2|\partial H|^2\left(M_P^2+2\xi |H|^2\right)}{\left[M_P^4+2\xi M_P^2|H|^2-2\alpha|\partial H|^2\right]}\,.{\label{CHISQUARE}}\ee

Again, we proceed introducing a background in the $\phi_1$ direction. Focusing on $\phi_1'\,\phi_2\,\rightarrow\,\phi_1'\,\phi_2$ scattering, we still have the relevant terms that we had in the Palatini-Einstein-frame analysis we made in the absence of the $R^2$ term but we also have the extra terms arising from  $\chi^2$ and $\chi^4$. Using ({\ref{CHISQUARE}}) and keeping relevant terms up to second order, we obtain
\be \chi^2\,\approx\,\frac{1}{M_P^2}\left((\partial\phi_1')^2+(\partial\phi_2)^2\,\right)\,.\ee
Similarly, for $\Omega^{-2}$ we have
\be \Omega^{-2}\,\approx\,\frac{M_P^2}{\xi\overline{\phi}_1^2}\left(1-2\frac{\phi_1'}{\overline{\phi}_1}+3\frac{{\phi_1'}^2}{\overline{\phi}_1^2}-\frac{\phi_2^2}{\overline{\phi}_1^2}-\frac{\alpha}{\xi M_P^2}\frac{(\partial\phi_1')^2}{\overline{\phi}_1^2}-\frac{\alpha}{\xi M_P^2}\frac{(\partial\phi_2)^2}{\overline{\phi}_1^2}\right)\,,\ee 
while, for $\Omega^{-4}\,\approx M_P^4/\xi^2\overline{\phi}_1^2$ the constant term suffices. Without much effort we can see that the only $\alpha$-dependent term that can contribute to scattering is
\be \frac{\alpha}{2\xi^2\overline{\phi}_1^4}(\partial\phi_1')^2(\partial\phi_2)^2\,=\,\frac{\alpha}{2M_P^4}(\partial\chi_1)^2(\partial\chi_2)^2\,,\ee 
written also in terms of the canonically normalised scalars
\be \chi_1\,=\,\frac{M_P}{\xi^{1/2}\overline{\phi}_1}\phi_1',\,\,\,\,\,\chi_2\,=\,\frac{M_P}{\xi^{1/2}\overline{\phi}_1}\phi_2\,.\ee 
The corresponding contact interaction quartic vertex is
\be V_{\chi_1^2\chi_2^2}\,=\,\frac{2i\alpha}{M_P^4}\left(p_{\chi_11}\cdot p_{\chi_12}\right)\left(p_{\chi_21}\cdot p_{\chi_22}\right)\,.\ee 
The new vertex modifies the result ({\ref{PAL}}) as
\be i{\cal{M}}\,=\,-\frac{iE^2}{\overline{\phi}^2}\left(8(1-\cos\theta)+\frac{2(1+\cos\theta)}{\xi(1-\cos\theta)}\right)+\frac{2i\alpha E^4}{\xi^2\overline{\phi}_1^4}(1-\cos\theta)^2\,,\ee 
where again $E$ is the center of mass energy in the Jordan frame and $\theta$ the angle between the incident and the outgoing $\phi_1$.

{\subsection{Metric Higgs Inflation in the presence of an $\bf { R^2}$ term}
We start again with the action ({\ref{ACTI}}) rewritten in terms of the auxiliary scalar $\chi$ as
\be {\cal{S}}\,=\,\int\,d^4x\,\sqrt{-g}\left\{\frac{1}{2}M_P^2\Omega^2R(g)\,-|\partial H|^2-\frac{\alpha}{4}\chi^4\right\}\ee 
in the framework of the metric formulation. Again, \be\Omega^2=1+\frac{1}{M_P^2}\left(2\xi|H|^2+\alpha\chi^2\right)\,.\ee
Going to the Einstein frame through $g_{ \mu\nu}\rightarrow\,\Omega^{-2}g_{ \mu\nu}$, we obtain
\be {\cal{S}}\,=\,\int\,d^4x\sqrt{-g}\left\{\frac{1}{2}M_P^2R-\frac{3M_P^2}{4\Omega^4}(\partial\Omega^2)^2-\frac{1}{\Omega^2}|\partial H|^2-\frac{\alpha\chi^4}{4\Omega^4}\right\}\ee
Introducing a background in the $\phi_1$-direction
\be \phi_1\,=\,\phi_1'\,+\,\overline{\phi}_1\,\,\,\,\,\,{\text{with}}\,\,\,\overline{\phi}_1>>M_P/\sqrt{\xi}\,,\ee
we have
\be \bear{l}
\Omega^{-2}\,\approx\,\frac{M_P^2}{\xi\overline{\phi}_1^2}\left(1-2\frac{\phi_1'}{\overline{\phi}_1}+3\frac{{\phi_1'}^2}{\overline{\phi}_1^2}-\frac{\phi_2^2}{\overline{\phi}_1^2}\right)\left(1+\frac{\alpha}{\xi}\frac{\chi^2}{\overline{\phi}_1^2}\right)^{-1}\\
\,\\
\Omega^{-4}\,\approx\,\frac{M_P^2}{\xi\overline{\phi}_1^2}\left(1-4\frac{\phi_1'}{\overline{\phi}_1}+10\frac{{\phi_1'}^2}{\overline{\phi}_1^2}-2\frac{\phi_2^2}{\overline{\phi}_1^2}\right)\left(1+\frac{\alpha}{\xi}\frac{\chi^2}{\overline{\phi}_1^2}\right)^{-2}
\eear\ee
The scalar $\chi$, initially introduced as an auxiliary, does propagate, having a non-canonical kinetic term. Introducing a canonical scalar in its place through
\be -\frac{1}{2}(\partial\rho)^2\,=\,-\frac{3\alpha^2M_P^2}{\xi^2\overline{\phi}_1^4}\frac{\chi^2(\partial\chi)^2}{\left(1+\frac{\alpha\chi^2}{\xi M_P^2}\right)^2}\,,\ee 
we obtain ($\rho(\chi=0)=0$)
\be \rho\,=\,\sqrt{\frac{3}{2}}M_P\ln\left(1+\frac{\alpha\chi^2}{\xi\overline{\phi}_1^2}\right)\,\,\,\,{\textit{or}}\,\,\,\,\,\,\chi^2\,=\,\frac{\xi}{\alpha}\overline{\phi}_1^2\left(e^{\sqrt{\frac{2}{3}}\frac{\rho}{M_P}}-1\right)\,.\ee
In a similar fashion we replace the non-canonical scalars $\phi_1'$ and $\phi_2$ with canonically normalised fields $\chi_1,\,\chi_2$ as
\be \phi_1'\,=\,\left(\frac{\xi}{1+6\xi}\right)^{1/2}\frac{\overline{\phi}_1}{M_P}\chi_1,\,\,\,\,\,\,\,\,\phi_2\,=\,\xi^{1/2}\frac{\overline{\phi}_1}{M_P}\chi_2\,.\ee

At this point we focus on the fact that we aim at calculating the scattering amplitude $\chi_1,\,\chi_2\,\rightarrow\,\chi_1,\,\chi_2$. Note however that part of this amplitude we have already calculated when we considered metric Higgs inflation in the Einstein frame in the absence of the $R^2$-term. Thus, what we really aim at here are the {\textit{extra $\alpha$-dependent contributions}}, which amount to just $\rho$-exchange in the $t$ channel. To compute this amplitude we need to analyse the interactions $\chi_1^2\rho$ and $\chi_2^2\rho$ that appear in various parts of the Lagrangian. Doing that we obtain
\be \bear{l}
{\cal{L}}_{\chi_1^2\rho}\,=\,\frac{(1+12\xi)}{\sqrt{6}(1+6\xi)M_P}\rho(\partial\chi_1)^2\,+\,\frac{18\xi}{\sqrt{6}(1+6\xi)M_P}\chi_1(\partial\chi_1\cdot\partial\rho)\\\,\\
{\cal{L}}_{\chi_2^2\rho}\,=\,\frac{1}{\sqrt{6}M_P}\rho(\partial\chi_2)^2\,-\frac{\sqrt{6}\xi}{M_P}\chi_2(\partial\chi_2\cdot\partial\rho)
\eear
\ee
These interaction terms lead to the vertices
\be \bear{l}
V_{\chi_1^2\rho}\,=\,-\frac{2i(1+12\xi)}{\sqrt{6}(1+6\xi)M_P}\left(p_{\chi_11}\cdot p_{\chi_12}\right)\,-\frac{18i\xi}{\sqrt{6}(1+6\xi)M_P}\left(p_{\chi_11}+p_{\chi_12}\right)\cdot p_{\rho}\\
\,\\
V_{\chi_2^2\rho}\,=\,-\frac{2i}{\sqrt{6}M_P}(p_{\chi_21}\cdot p_{\chi_22})\,+\,\frac{i\sqrt{6}\xi}{M_P}\left(p_{\chi_21}+p_{\chi_22}\right)\cdot p_{ \rho}
\eear\ee
Apart from the $\rho$-interaction vertices we also need its propagator. It turns out the $\rho$ is massive, having a mass 
\be m_{\rho}^2\,=\,\frac{M_P^2}{3\alpha}\,\,\,\,\Longrightarrow\,\,\,\frac{i}{p^2-m_{\rho}^2}\,.\ee
Combining the above we arrive at the corresponding {{$\alpha$-dependent part of the amplitude}}
 \be i{\cal{M}}_{\rho,t}^{(\alpha)}\,=\,\frac{i\alpha(1-6\xi)}{2M_P^4}\frac{t^2}{1-\frac{3\alpha t}{M_P^2}}\,.\ee
In the Einstein-frame center of mass energy $\tilde{E}$ we have $t=-2\tilde{E}^2(1-\cos\theta)$, in terms of the angle $\theta$ between incoming and outgoing $\chi_1$. On the other hand, in terms of the Jordan frame energy $E^2\,\approx\,\frac{\xi\overline{\phi}_1^2}{M_P^2}\tilde{E}^2$, we have 
\be i{\cal{M}}_{\rho,t}^{(\alpha)}\,=\,\frac{2i\alpha(1-6\xi)}{\xi^2}\left(\frac{E}{\overline{\phi}_1}\right)^4\frac{(1-\cos\theta)^2}{\left(1+\frac{6\alpha}{\xi}\frac{E^2}{\overline{\phi}_1^2}(1-\cos\theta)\right)}\,.\ee

\section{Amplitudes in the electroweak vacuum}

In a recent article \cite{M}, it has been argued that, in the Palatini formulation, the naive unitarity violation scale $M_U = M_P/\sqrt{\xi}$ in the electroweak vacuum is flawed, and a claim has been made that the correct unitarity violation scale can be deduced from the amplitude in the presence of a large background. The argument is based on an estimation of the magnitude of the fields $\phi_v = \langle\phi^2\rangle^{1/2} \sim E$ in the interaction volume. The Palatini formalism action in the Einstein frame is:
\be {\cal{S}}\,=\,\int\,d^4x\sqrt{-g}\left\{\frac{1}{2}M_P^2R-\frac{1}{\Omega^2}|\partial H|^2\,\right\}\, \hspace{\baselineskip}\text{with}\hspace{\baselineskip}\Omega^2=1+\frac{2\xi|H|^2}{M_p^2}\,.\ee 
Here, the unitarity unitarity violation scale $\Lambda$ is deduced by developing $\Omega^{-2}|\partial H|^2$ to obtain for instance:
\begin{equation}
\Omega^{-2}|\partial H|^2 \supset \frac{\xi}{2M_P^2}\phi_2^2(\partial\phi_1)^2\,,
\end{equation}
from where we read $\Lambda \sim M_p/\sqrt{\xi}$. But with the estimate $\phi_{1, 2} \sim E$ the expansion of $\Omega^{-2}$ starts to fail precisely when the energy approaches $M_p/\sqrt{\xi}$, calling for a different way to obtain the cutoff $\Lambda$. 

In the proposal of~\cite{M}, one uses the estimate for the magnitude of fields $\phi_v \sim E$ and the assumption that a background $\bar\phi \ll E$ shall not affect the amplitude, which can in this case approximate the one in the electroweak vacuum. The amplitude computed in the presence of a background $\bar\phi \gg E$ is then matched with the one in the electroweak vacuum at $\bar\phi \approx E$, in some kind of threshold approximation. If this is right, we can take the limit $\bar\phi_1 \rightarrow E$ in equation (\ref{PAL}) as the amplitude in the electroweak vacuum:
\begin{equation}
i\mathcal{M}^{(P)}_{\bar\phi_1=0} = -8i(1-\cos\theta) - \frac{2i(1+\cos\theta)}{\xi(1-\cos\theta)}\,.
\end{equation}
In the limit where $\xi$ is large (and $\cos\theta\neq 1$), only the first term remains. Most importantly, this amplitude does not grow with energy, suggesting that, besides a marginal degree of violation, unitarity is preserved up to the Planck scale.

Now, in the metric formulation, there are additional terms in the action (\ref{action_metric_einstein}), such as:
\be {\cal{L}}_{12}\,=\,-\frac{6\xi^2\phi_1\phi_2}{M_P^2\Omega^4}(\partial\phi_1\cdot\partial\phi_2)\,\ee
which, if $\Omega^4 \simeq 1$, induces a scale of unitarity violation $\Lambda \sim M_p/\xi$. Indeed, using again $\phi_{1, 2} \sim E$, the approximation $\Omega^4 \simeq 1$ is valid up to that scale in the large $\xi$ limit. So contrary to the case of the Palatini formulation, this estimation for the cutoff $\Lambda$ seems consistent and we may accept it.

However, one can apply the same reasoning as before, in order to obtain the amplitude in the electroweak vacuum from the one in the presence of a large background.\footnote{One should take however into account that the field $\phi_1$ is not canonically normalised in the Jordan frame in the presence of a large background, as we have seen in (\ref{CANO}), spoiling the estimate of the magnitude of the fields $\phi_v = \langle\phi^2\rangle^{1/2} \sim E$ on which the argument of \cite{M} is based on.} Doing so, we take the limit $\bar\phi_1 \rightarrow E$ in equation (\ref{JMA}) as an approximation for the amplitude in the electroweak vacuum:
\be i\mathcal M_{\bar\phi_1=0}^{(M)} = -\frac{8i(1+3\xi)}{1+6\xi}(1-\cos\theta)-\frac{2i(1+\cos\theta)}{\xi(1-\cos\theta)}\,.\ee
Just as in the Palatini case, the amplitude does not grow with energy, suggesting that unitarity is preserved up to high scale in the electroweak vacuum.

This contradicts the cutoff $\Lambda \sim M_p/\xi$ that was obtained previously. So either this cutoff is too naive, or the argument to connect amplitudes in the presence of a background to amplitudes in the electroweak vacuum is not correct. To test this assumption, let us use it for the following toy Lagrangian:
\begin{equation}
\mathcal{L} = -\frac{1}{2}(\partial\phi_1)^2 - \frac{1}{2}(\partial\phi_2)^2 - \frac{\alpha M}{2}\phi_1\phi_2^2\,.
\end{equation}
In the ``electroweak vacuum" $\bar\phi_1=0$, the amplitude for $\phi_1\phi_2 \rightarrow \phi_1\phi_2$ scattering is: 
\begin{equation}
i\mathcal{M}_{\bar\phi_1=0}^{\mathrm{exact}} = -\frac{i\alpha^2M^2}{4E^2} +\frac{i\alpha^2M^2}{2E^2(1+\cos\theta)}
\end{equation}
with $E$ the center of mass energy and $\theta$ the angle between incoming and outgoing $\phi_1$.
We may now expand $\phi_1$ around an arbitrary background $\bar\phi_1 \neq 0$. This induces a mass to $\phi_2$, and the amplitude becomes:
\begin{equation}
i\mathcal{M} = -\frac{i\alpha^2M^2}{4E^2-\alpha M\bar\phi} +\frac{i\alpha^2M^2}{2E^2(1+\cos\theta)+\alpha M\bar\phi}\,.
\end{equation}
Of course, this gives back $\mathcal{M}_{\bar\phi_1=0}^{\mathrm{exact}}$ when $\bar\phi_1=0$. If the argument described above is correct, we can approximate the amplitude $\mathcal{M}_{\bar\phi_1=0}^{\mathrm{exact}}$ with the amplitude when $\bar\phi \rightarrow E$ in this last equation. However, this is clearly wrong as soon as $E \ll \alpha M$. Therefore, the validity of the argument in the more complicated case of Higgs inflation is questionable.

\section{Conclusions}

In this work, we analysed the ultraviolet behaviour of Higgs inflation models as well as models with an $R^2$ term, in both the metric and Palatini formulations of gravity. {{Although the presence of a large non-minimal coupling introduces interactions that violate tree-level unitarity at a scale much lower than the Planck scale within the standard electroweak vacuum, we show that this is not the case in the presence of a large inflaton background. Indeed all tree-level 4-point scattering amplitudes remain small on-shell at high energies, lower than the background, which plays the role of the effective ultraviolet cutoff of the theory, taking values up to the Planck scale. Our conclusions are based on the gauge-invariant and frame-invariant computation of Higgs scattering amplitudes in the framework of both the standard metric formulation of gravity and the Palatini formulation.}}

\section*{Acknowledgments}
Work partially performed by I.A. as International professor of the Francqui Foundation, Belgium.

\end{document}